\documentclass[a4paper,11pt]{article}

\usepackage{lineno,hyperref}
\usepackage{tikz}
\usepackage{tikz-cd}
\usepackage{dsfont}
\usepackage{amsfonts}
\usepackage{amsmath}
\usepackage{slashed}
\usepackage{graphics}
\usepackage{amssymb}
\usepackage[compat=1.1.0]{tikz-feynman}
\usepackage{setspace}
\usepackage{float}
\usepackage{cleveref}
\usepackage{units}
\usepackage{amscd}
\usepackage[left=1in, right=1in, top=1in, bottom=1in]{geometry}
\usepackage{nicefrac}
\usepackage{subcaption}
\usepackage{changepage}
\usepackage{setspace}

\modulolinenumbers[5]

\providecommand{\keywords}[1]{%
  \vspace{0.5cm}
  \noindent\textbf{Keywords: } #1
  \vspace{0.5cm}
}

\title{\textbf{Testing Quantum Gravity with Gravitational Waves from the Ringdown of Binary Black Hole Coalescences: A New Frontier in Fundamental Physics}}

\author{
  M.D.C.~Torri$^{12}$\thanks{Corresponding author: marco.torri@unimi.it, marco.torri@mi.infn.it}\;,
  F.~Ricci$^{34}$, M.~Giammarchi$^{12}$, L.~Miramonti$^{12}$, V.~Toso$^{12}$, C.~Sigala$^{1}$\\
  \\
  \textit{$^1$Dipartimento di Fisica, Università degli Studi di Milano,} \\
  \textit{via Celoria 16, I-20133 Milano, Italy} \\
  \textit{$^2$INFN Milano,} \\
  \textit{via Celoria 16, I-20133 Milano, Italy} \\
  \textit{$^3$Dipartimento di Fisica, Sapienza Università di Roma,} \\
  \textit{Piazza Aldo Moro 5, I-00185 Roma, Italy} \\
  \textit{$^4$INFN Roma,} \\
  \textit{Piazza Aldo Moro 5, I-00185 Roma, Italy}
}

\begin{document}

\maketitle


\abstract{The observation of gravitational waves emitted during the merging phase of compact binary coalescing objects has opened a new field of investigation in fundamental physics. It is now possible to test the predictions of General Relativity with unprecedented precision in the strong gravitational field regime. These initial observations therefore call for further research, as the detection of gravitational waves emitted by coalescing black holes may allow the investigation of the properties of spacetime near the event horizon, also providing valuable information on the structure of these objects. This also opens the possibility of testing predictions from quantum gravity models regarding the presumed quantized structure of black holes, related to the quantization of their surface and, consequently, their entropy. In the future, the considerable amount of data obtained by the LIGO-Virgo-KAGRA collaboration will be followed by observations from next-generation interferometers such as the Einstein Telescope or the Cosmic Explorer. It is therefore of great interest to explore the potential of gravitational wave observations for investigating aspects of quantum gravity, which we will address considering the special case of the ringdown emission following the coalescence of binary black hole systems.}

\keywords{gravitational waves, ringdown, black hole, quantum gravity}


\section{Introduction}
\label{sec:introduction}

The first direct observation of a ringdown emission event from the coalescence of two black holes (BHs), featuring different overtones of quasi-normal modes (QNMs), represents a significant step toward using gravitational waves (GWs) to probe the structure of BHs \cite{KAGRA}. In fact, in the classical scenario, the emission frequencies of GWs during the ringdown are expected to have a discrete spectrum. The experimental confirmation of different frequencies in QNM emission imposes constraints on models describing BHs. This is particularly relevant for scenarios proposing the quantization of BHs structure.

In many theoretical frameworks, black hole (BH) quantization is rooted in the Bekenstein–Hawking relation \cite{Bekenstein, Bekenstein2, Meissner, Sahlmann, Maldacena,Bekensteinbound, Berti2, Carneiro}, which links entropy to the event horizon (EH) area. The correspondence between entropy, area, and consequently BH mass has motivated several approaches. Indeed, the hypothesis of black hole area quantization (BHAQ) is well established in several models, including string theory (ST) and loop quantum gravity (LQG). In the Bekenstein–Mukhanov model \cite{Bekenstein3}, quantization arises from a discrete spectrum of the EH surface—expressed in units of the Planck area, $l_P^2$—where macroscopic levels correspond to quantized BH energy levels (mass eigenstates), in close analogy with atoms in the Bohr model \cite{Hod}. The Bekenstein-Mukhnanov model does not contemplate a further fine subdivision of the energy levels. In contrast, in ST the entropy is obtained via the Cardy formula by counting excited brane microstates \cite{Strominger}, while in LQG it arises from combinatorial counting of punctures in the spin network on the EH \cite{Meissner,Sahlmann}. In these two cases, entropy is tied to macroscopic levels associated with the BH surface, but with a large degeneracy arising from multiple microscopic configurations. In all these scenarios, the entropy quantization implies the BHAQ and thus the quantization of its mass. Accordingly, GW emission during ringdown would occur through quantized transitions between different mass-energy levels. Moreover, the ringdown emission must remain compatible with the discrete structure of the classically predicted QNM spectrum.

The ringdown emission phase of GW is of particular interest, as it encodes direct information about the remnant BH. GW emission from coalescing objects proceeds through three different phases: inspiral, merger and ringdown. After the merging, during the ringdown, the newly formed BH radiates energy in the form of GWs. The ringdown signal can be modeled as a superposition of damped armonic oscillators describing the relaxation of the BH to equilibrium, leading to a decrease in energy-mass. Searching for quantum signatures therefore requires identifying deviations in the predicted ringdown frequencies that could arise from the quantum structure of BH mass eigenstates.

Detecting and analyzing QNMs from ringdown thus may provide a rare opportunity to probe the interface between General Relativity (GR) and Quantum Mechanics (QM), whose unification remains an open challenge \cite{Cost,WP}. In this work, we explore the potential of using ringdown GWs to extract information about the presumed quantum structure of BHs and to constrain possible quantization scenarios of spacetime. By exploiting the ringdown, one can look for modifications induced by the BHAQ hypothesis both far from the BH and near its surface, in particular between the EH and the photon sphere. If BHs possess an underlying quantized structure, such as foreseen in BHAQ scenario, the emitted GWs may carry imprints of this quantization \cite{Foit}. Indeed, the BHAQ hypothesis can introduce changes in the discrete frequencies predicted for the QNMs as detected by distant observers. This appears to be the most promising framework to detect the presumed signatures of BH quantization. However, the BHAQ scenario can also produce echo effects in the propagation of GWs, caused by modifications related to the geometry close to the BH, that is, between the EH and the photon sphere \cite{Foit}. This second hypothesis seems to be disfavored by the difficulties related to the detection of GWs during the fast and cahotic phase of ringdown emission \cite{Coates}.

The scientific relevance of this research is reinforced by the growing number of GW detections from the LIGO-Virgo-KAGRA network, which is delivering an unprecedented dataset of statistical and detailed studies of BH ringdowns. The next generation of detectors, such as the Eisntein Telescope (ET) \cite{ET_2} and the Cosmic Explorer (CE) \cite{Cosmic_Explorer}, will feature such an increase in sensitivity that it will have the potential to carry out this research. With substantially higher sensitivity over a broader frequency range and leveraging the larger dimension of the interferometers, they will resolve ringdown modes with greater precision, and potentially uncover subtle features indicative of BH quantization or deviations from classical predictions.

In this work we will introduce the equations governing GW emission for static and rotating sources in the frameworks developed by Regge–Wheeler and Zerilli (Schwarzschild spacetime), and Teukolsky (Kerr-Newman spacetime). We will then show how the hypothesis of black hole quantization can be incorporated, considering the relation between entropy, area, and mass. Next, we will examine various quantization scenarios and their corresponding effects on GW emission during the ringdown phase. We will illustrate how quantization hypotheses have to remain compatible with the experimental observation of different QNM discrete frequencies. We will also discuss the possible impact of certain quantum gravity (QG) models on GW propagation. Finally, we will consider how these hypotheses could be tested with the aid of next-generation interferometers.

\section{Quasi Normal Modes and the Ringdown emission phase}
\label{sec:QNMemission}

After the merger, the newly formed compact object, usually a BH, undergoes a settling process known as the ringdown phase \cite{Damour}. The emission phase encodes information on the structure of BHs \cite{Berti,Nollert}, as it describes GW emission from the relaxation of the final BH after merger. During the ringdown, the object emits GWs as it dissipates residual oscillations. These waves have characteristic frequencies and decay exponentially over time as the BH approaches a stable, stationary state.

Using the linear approximation, it can be shown that the perturbations associated with the ringdown are described by Schr\"odinger equations, expressed using the tortoise radial coordinate $r_{*}$.\footnote{The tortoise coordinate is defined as: $\frac{dr_{*}}{dr}=\frac{r^{2}+a^{2}}{\Delta}$, where $\Delta=r^2-\frac{2GM}{c^2}r+a^2+Q_g^2$, $Q_g=\sqrt{\frac{G}{4\pi}}\frac{Q}{c^2}$, $Q$ is the electric charge, and $a=j/Mc$ is the spin parameter for a generic Kerr-Newman BH.}

For Schwarzschild BHs, axial perturbations with odd parity are described by the Regge–Wheeler function $\Psi_{RW}$ satisfying \cite{Regge}:
\begin{equation}
\label{Schr}
\frac{d^{2}\Psi_{RW}(r_{*})}{dr_{*}^2}+\left[\omega^{2}-V_{RW}(r_{*})\right]\Psi_{RW}(r_{*})=0,
\end{equation}
with associated effective potential:
\begin{equation}
\label{effpot1}
V_{RW}(r_{*})=f(r_{*})\left(\frac{l(l+1)}{r_{*}^{2}}-\frac{6GM}{c^2\,r_{*}^{3}}\right).
\end{equation}
On the other hand, even-parity perturbations for the Schwarzschild BH are governed by the Zerilli function satisfying the same equation \cite{Zerilli,Moncrief,Martel} with effective potential defined as:
\begin{align}
\label{effpot2}
&V_{Z}(r_{*})=\frac{2f(r_{*})}{r_{*}^{3}(\Lambda r_{*}+3M)^{2}}\left[\Lambda^{2}(\Lambda+1)r_{*}^{3}+3\,\frac{GM}{c^2}\Lambda^{2}r_{*}^{2}+9\left(\frac{GM}{c^2}\right)^{2}\Lambda r_{*}+9\left(\frac{GM}{c^2}\right)^{3}\right],\\
&f(r_{*})=1-\frac{2GM}{c^{2}\,r_{*}}, \qquad
\Lambda=\tfrac{1}{2}(l-1)(l+2). \notag
\end{align}
For rotating or charged BHs, the Teukolsky equation \cite{Teukolsky,Arbey} after the Sasaki–Nakamura transformation \cite{Sasaki,Sasaki2} reduces to \cref{Schr} with effective potential:
\begin{align}
\label{effpot3}
&V_{SN}(r_{*})\simeq\frac{\Delta}{(r_{*}^2 + a^2)^2} \left[ \lambda + \frac{\Delta'}{r_{*}^{2}+a^{2}}r_{*}-\frac{3\Delta r_{*}^2}{(r_{*}^2 + a^2)^2}  \right],\\
&\Delta=r_{*}^2-\frac{2GM}{c^{2}}r_{*}+a^2 \notag
\end{align}
where $a$ is the spin parameter defined before, $\Delta'(r)=d\Delta/dr$, and $\lambda_{lm}$ encodes the $l,\,m$ dependence.

Asymptotically, solutions behave as plane waves:
\begin{equation}
\psi(r_{*})\sim e^{+i\omega r_{*}}\quad r_{*}\rightarrow+\infty,\qquad\psi(r_{*})\sim e^{-i\omega r_{*}}\quad r_{*}\rightarrow-\infty,
\end{equation}
corresponding to outgoing waves at infinity and ingoing waves at the horizon. This boundary condition turns the perturbation problem into a non-Hermitian eigenvalue problem. Only specific complex frequencies satisfy both conditions simultaneously, giving rise to the quasi-normal modes (QNM).\\
The main analytic approaches are WKB, Pöschl–Teller, and the continued fraction method \cite{Schutz,Iyer,Leaver,Ferrari}, with the latter being the most accurate for Kerr and Kerr–Newman BHs. In the WKB framework \cite{Schutz,Iyer} one can find an approximate analytic form for the allowed frequencies:
\begin{equation}
\omega_{nlm}\simeq\sqrt{V_{0}}\pm i\left(n+\frac{1}{2}\right)\sqrt{2|V_{0}''|},
\end{equation}
Assuming the asymptotic behaviour of the effective potential $V_{0}\simeq \frac{l(l+1)}{27M^2}$, the frequency becomes:
\begin{equation}
\omega_{nlm}\simeq\frac{c^{3}}{GM}\left(\frac{\ln{3}}{8\pi M}+\frac{1}{4M}i\left(n+\frac{1}{2}\right)+O(\sqrt{n})\right), 
\end{equation}
hence the frequency presents its trend $\omega\sim1/M$ and the perturbation can be written as:
\begin{equation}
h_{\mu\nu}^{nlm}(t,\,x)=\sum_{n,\,l}A_{\mu\nu}^{\;\;lnm}e^{\left(-\mathfrak{Im}(\omega_{nlm})(t-t_{0})\right)}\cos{\left(\mathfrak{R}(\omega_{nlm})(t-t_{0})\right)},
\end{equation}
where the polarization tensor $A_{\mu\nu}$ encodes the wave amplitude.

\section{Black Holes and Quantum Mechanics}
\label{sec:BHQM}

BHs may be inherently quantum systems, since their surface is linked to entropy through the Bekenstein-Hawking relation \cite{Bekenstein, Bekenstein2,Meissner,Sahlmann,Berti2,Carneiro,Maldacena}:
\begin{equation}
\label{entropy}
S=k_B \frac{A}{4\,l_{Pl}^2}
\end{equation}
and their quantum instability may manifest through the Hawking radiation \cite{Hawking3,Hawking4}.
The ability to relate the entropy of a BH to its surface area represents a fundamental result that can be used to address the challenge of unifying GR and QM. The possibility of encoding all quantum information on the surface of a spacetime region, as happens in the case of a BH and the information paradox \cite{Polchinski}, lies at the foundation of the holographic principle \cite{Maldacena1}. This deep correspondence between area, entropy and energy has inspired different quantization scenarios, such as the Bekenstein-Mukhanov proposal \cite{Bekenstein3}, ST \cite{Strominger} and LQG \cite{Meissner,Sahlmann}. In all these frameworks, entropy is quantized, which implies a discrete EH area and a quantization of the BH mass via the relation $A\sim M^2$.

A consequence of all the QG models considered is that the BH surface is quantized and can be considered in the form:
\begin{equation}
\label{a1}
A=4\,\frac{l_{Pl}^{2}}{k_B}\,S=\alpha\,l_{Pl}^{2}\overline{n},
\end{equation}
where $\overline{n}$ is a quantization index representing the number of fundamental surface elements, and $\alpha$ is an appropriate proportionality coefficient, depending on the considered model, so it is possible to derive the equation \cite{Foit}:
\begin{equation}
\label{a2}
\Delta A=\alpha\,l_{Pl}^{2}\,\Delta\overline{n}=\alpha \frac{\hbar G}{c^{3}}\,\Delta\overline{n}
\end{equation}
Introducing the dependence of the area on the Schwarzschild radius $r=2GM/c^2$, it is possible to obtain the following relation between $\Delta A$ and $\Delta M$:
\begin{equation}
\label{d1}
\Delta A\simeq\frac{32\pi\,G^{2}}{c^{4}}M\Delta M
\end{equation}
valid in first approximation for any BH spacetime. From the area-mass relation, one finds that each quantized mass transition produces an emitted energy $\hbar\,\omega_{QNM}\sim\hbar/M$. In Bekenstein "black hole atom" picture \cite{Bekenstein3,Hod}, this transition corresponds to the emission of a single graviton which carries away the energy of the transition, where $\Delta E_{M}=E_{M}^{(\overline{n}+1)}-E_{M}^{(\overline{n})}=(M^{(\overline{n}+1)}-M^{(\overline{n})})c^{2}$ is the energy jump between two closely spaced energy-mass eigenstates (denoted by $\overline{n}$ and $\overline{n}+1$) \cite{Foit}. However this scenario seems to be incompatible with the observed QNM spectrum. Experimental observations now provide complementary input, indeed, GW detectors have confirmed the existence of QNMs in the post-merger ringdown, exhibiting a spectrum of different discrete frequencies. Therefore any consistent quantization scheme must be in accordance with this constraint. In fact, at a macroscopic level a BH must produce a set of sharp spectral lines during the ringdown GW emission phase. For instance, in a binary BH merger the dominant $l=2, n=0$ QNM radiates an energy of order $\Delta E_{\rm QNM}\sim10^{-3} M c^{2}$, i.e. $\sim10^{63}\,\mathrm{eV}$ for a remnant of $100\,M_{\odot}$. In the Bekenstein picture, the elementary jump is:
\begin{equation}
\Delta E_{\rm Bek}\sim\frac{\hbar c^{3}}{GM},
\end{equation}
which, for a $100 M_{\odot}$ remnant BH corresponds to $\sim10^{-12}\,\mathrm{eV}$. A cascade of
\begin{equation}
N_{\text{tot}}=\frac{\Delta E_{\rm QNM}}{\Delta E_{\rm Bek}}=10^{-3}\frac{GM^2}{\hbar\,c}
\end{equation}
total emitted gravitons is foreseen, that is a total number of $N_{\rm tot}=\sim10^{75}$ quantized energy jumps for every QNM mode emitted, for a $100\,M_{\odot}$ BH remnant. On the other hand, the ratio between the fundamental energy emitted by a single graviton as foreseen in the Bekenstein quantized scenario, $\omega_0$, and the classical QNM energy is:
\begin{equation}
\frac{\hbar\,\omega_0}{\hbar\,\Omega_{\text{QNM}}} = \frac{\omega_0}{\Omega_{\text{QNM}}} = \frac{\alpha}{32\,\pi\,k}\,\Delta\overline{n}.
\end{equation}
For a typical value $k=0.3737$ (for the fundamental $l=2$ QNM), and $\alpha=4\log{3}$, as in the Bekenstein–Mukhanov model, or $\alpha$ equal to the Barbero–Immirzi parameter in LQG, the ratio becomes $\omega_0/\omega_{QNM}\sim (12\times\Delta\overline{n})\%$. Hence, for each classical QNM frequency, fewer than about ten Bekenstein jumps are possible, not enough gravitons to carry away all the emitted energy. This discrepancy disfavors the Bekenstein picture of single-graviton emission, in favor of a scenario involving the coherent and collective emission of many quantum states for each quantized mass jump.

In the Bekenstein scenario, BH entropy is quantized in levels corresponding to different area eigenstates via \cref{entropy}. By contrast, in ST and LQG the entropy is realized as an exponential degeneracy of microstates, with the total number of configurations $\mathcal{N}\simeq e^{S}$ for given entropy value $S$. Therefore, for every macrolevel of entropy a large number of microstates with similar entropy is predicted. Thanks to the correspondence between entropy and mass, this naturally leads to the hypothesis that the energy spectrum is not perfectly sharp but consisting of a large number of degenerate levels with similar energies, concentrated around the energy levels corresponding to the different area eigenstates. The microscopic level spacing inside each band is $\delta E_{micro}\sim \delta E_{M}\,e^{-S}$, where it is supposed to be $\delta E_{M}<\Delta E_{M}$. 

The introduction of this degeneracy can provide the required number of level jumps, corresponding to the emission of a large number of gravitons--each with similar energy--capable of carrying all the emitted energy as the sum of the contributions of the individual gravitons. Without a complete QG theory at hands, to study how such degeneracy affects the BH energy configurations, we model its ringdown emission semiclassically using the Lee–Friedrichs formalism \cite{Xiao,Lonigro}. By coupling a bound state to an environment with many degrees of freedom, it explains how discrete states may acquire finite lifetimes and broadened energy spectra, providing a simplified effective framework for studying unstable quantum systems. This model describes a discrete quantum state interacting with a continuum of states, leading to a decay rate and resonance phenomena that represent the continuous quantum vibration modes $|\omega\rangle$ of the BH during GW emission, coupled through self-interaction, which accounts for the backreaction on the BH geometry:
\begin{align}
\label{LF}
\mathcal{H}&=\mathcal{H}_{GW}+\mathcal{H}_{Diss}+V_{QNM}= \notag\\
&=\int_{0}^{+\infty} \hbar\omega|\omega\rangle\langle\omega|d\omega+\int_{0}^{+\infty}\frac{-i\hbar}{2} \gamma(\omega)|\omega\rangle\langle\omega|d\omega+\int_{0}^{+\infty}\left[g(\omega,\,\omega')|\omega\rangle\langle\omega'|+\text{h.c.}\right]d\omega\,d\omega'
\end{align}
where the first term $\mathcal{H}_{GW}$ describe different BH modes, the second term $\mathcal{H}_{Diss}$ encodes the dissipation present in the GW emission, and $V$ represents the modes self interaction, i.e. the emitted gravitons backreaction on the BH. The experimental QNM detection of different discrete frequencies indicates that the BH is well described by linear perturbations, hence the spacetime deviates only slightly from the final equilibrium state. Our model therefore provides a valid first-order approximation, assuming the BH is close to equilibrium and almost stationary. In fact, the ringdown GW emission phase is very fast and the resulting BH mass variation is about only $10^{-2}-10^{-3}$ of its initial value. 

Strictly speaking, QNMs are not normalizable in the conventional Hilbert space; indeed, they are solutions of a dissipative system. Therefore, the effective Hamiltonian defined here is non-Hermitian, as it must include the dissipative term. In order to highlight the QNM discrete frequencies, we introduce the resonant emission state in a band centered on the foreseen frequencies $\Omega^{n}_{M}$, emitted for every mass eigenstate jump, defined as:
\begin{equation}
|\Psi_{n}\rangle=\int_{\Omega^{n}_{M}-\omega_n/2}^{\Omega^{n}_{M}+\omega_n/2} c_{n}(\omega)|\omega\rangle d\omega,
\end{equation}
where $\omega_n$ is the $n-$th band amplitude. Requiring the normalization of the state:
\begin{equation}
\int_{\Omega^{n}_{M}-\omega_n/2}^{\Omega^{n}_{M}+\omega_n/2}|c_{n}(\omega)|^2d\omega=1,
\end{equation}
this leads to the condition:
\begin{equation}
c_{n}(\omega)\sim \frac{1}{\sqrt{\omega_n}}.
\end{equation}
The coherent emission of GW can be written as a linear combination of the coherent states $\sum_{n}\alpha_{n}|\Psi_{n}\rangle$.
Using the Feshbach formalism and projecting the Hamiltonian \cref{LF} onto the subspace spanned by all the combinations of the coherent states $|\Psi_n\rangle$ with the projector $P$, one obtains the Hamiltonian:
\begin{equation}
\mathcal{H}_{P}=P\mathcal{H}P+P\mathcal{Q}\frac{1}{E-Q\mathcal{H}Q}Q\mathcal{H}P=\sum_{n}\alpha_{n}\bigg(\hbar\,\Omega^{n}_{M}+\Theta_n(\omega)\bigg)|\Psi_{n}\rangle\langle\Psi_{n}|
\end{equation}
where $Q\perp P$ is the projector on the orthogonal subspace, the sum is made on the set of discrete frequencies, and $\Theta_n(\omega)$ is the self-energy introduced by the coupling to the continuum of every coherent state $|\Psi_{n}\rangle$:
\begin{equation}
\Theta_{n}(\omega)=\int\frac{|G_n(\omega')|^{2}}{\omega-\omega'+i0^{+}}d\omega'
\end{equation}
with the coupling integral defined as:
\begin{equation}
G_n(\omega)=\int_{\Omega^{n}_{M}-\omega_n/2}^{\Omega^{n}_{M}+\omega_n/2}g(\omega,\,\omega')c_n(\omega')d\omega'.
\end{equation}
The resonance amplitudes related to the discrete QNM emission can be computed using the Sokhotski-Plemelj identity:
\begin{equation}
\Gamma_{n}(\omega)=2\pi\rho(\omega)|G_n(\omega)|^{2}.    
\end{equation}
Since the density of states behaves as $\rho(\omega)\sim e^{S}$, the Eigenstate Thermalization Hypothesis (ETH)  model suggests to constraint the coupling kernel using the thermal coherence scale $g(\omega,\,\omega')\sim g_{0}e^{-S/2}$ \cite{ETH}. For a narrow and nearly flat packet, the coupling integral scales as $G_{n}(\omega)\sim g_{0}e^{-S/2}\sqrt{\omega_{n}}$. In this limit the resonance amplitudes can be written as:
\begin{equation}
\Gamma_{n}(\omega)\sim g_0^2\, \hbar\omega_{n},
\end{equation}
using the definition of the density of states $\rho$ and the coupling kernel $g$. For the resonance to remain sharply peaked--consistent with the discrete QNM frequencies experimentally observed--we must demand that the contribution of the band remains much smaller than the characteristic QNM energy scale:
\begin{equation}
\omega_{n}\ll \Omega_{M},
\end{equation}
feature required for coherent emission and coherent propagation for cosmological distances of GW. In this case, each classical QNM observed in GW detectors corresponds to a highly occupied coherent state, represented as a linear combination of the states associated with the jumps between area eigenstates, consistent with the macroscopic energy carried by the mode. Therefore, a coherent superposition of harmonic components with appropriate amplitudes and phases can produce effective oscillations with discrete frequencies, similar to those classically predicted for the QNMs. The emission should proceed through the release of a coherent GW packet at each energy transition, meaning that the presumed gravitons with degenerate micro–energy levels are expected to be emitted almost simultaneously at every quantized Bekenstein energy jump. Due to the short duration of the ringdown, BHs must emit GWs as a superposition of coherent graviton packets very rapidly. It can therefore be hypothesized that they behave similarly to a Bose-Einstein condensate quantum superfluid \cite{Manikandan,Hennigar}, emitting collective modes through a mechanism somewhat analogous to Dicke superradiance. Moreover, the idea of collective and coherent graviton emission resembles the scenario proposed in \cite{Dvali,Dvali2}, where BHs are conceived as condensates of a very large number of coherent quantum states--i.e. gravitons.

The implication is that the Bekenstein hypothesis can be consistent with QNM discrete spectrum by including the microdegeneration of entropy and energy as foreseen in ST or LQG. In theories with microstate degeneracy, emission may proceed through a coherent superposition of many modes whose collective contribution reproduces sharp QNM frequencies. In the strongly coherent limit, with the presumed emission of a large number of collective and coherent gravitons, favored by the first experimental evidences \cite{KAGRA}, observable out come is a discrete set of ringdown lines, possibly shifted relative to the classical prediction by the introduction of quantization of the BH area.

\section{Quantum Gravity in the ringdown GW emission}
\label{sec:emission}

During the ringdown phase, a BH is expected to emit GWs at the expense of its internal energy, thereby reducing its mass–energy. Under the BHQA hypothesis, GW emission may occur at discrete frequencies corresponding to transitions between quantized energy levels. The behavior of the BH is predicted to resemble that of a quantum system, with energy emitted proportional to the quantized energy of the waves \cite{Hod}. As illustrated before, the BH mass–energy should be quantized, since its surface area is related to its mass as $A\sim M^{2}$. In the case of a rotating and charged Kerr-Newman BH, the area can be written as:
\begin{equation}
A=\frac{8\pi M^{2}G^{2}}{c^{4}}\left[1+\sqrt{1-\frac{J^2c^2}{G^2M^4}-\frac{Q^2}{4\pi\epsilon_{0}\,GM^2}}\right]-\frac{GQ^2}{\epsilon_0\,c^4},
\end{equation}
where $J$ is the angular momentum, $a$ is the spin parameter, and $Q$ is the electric charge. The mass increment becomes:
\begin{equation}
\label{dA}
\Delta A=\frac{16\pi\,G^2M}{c^{4}}\left(1+\frac{1}{\sqrt{1-a^2}}\right)\Delta M-\frac{8\pi\,Ga}{c^{4}\,\sqrt{1-a^2}}\Delta J+\mathcal{O}(Q)
\end{equation}

In a damped oscillator described by the Schrödinger-type \cref{Schr} equation, the ringdown energy depends on the allowed frequencies, which correspond to the poles in the complex plane of the Fourier transform of the Green function of \cref{Schr}:
\begin{equation}
\left[\frac{d^{2}}{dr_{*}^{2}}+\left(\omega^{2}_{\overline{n}lm}-V_{eff}(r)\right)\right]G(\omega,\,r_{*},\,r_{*}')=\delta(r{*}-r_{*}'),
\end{equation}
where we again use the notation $\overline{n}$ to indicate that this index is related to the BHAQ scenario.
As a result, the supposed quantized energy of the GW perturbation related to the mass gap and emitted during the ringdown can be written as \cite{Foit,Maggiorequant}:
\begin{equation}
\label{d2}
\Delta M\,c^{2}=\hbar\,|\omega_{\overline{n}lm}|=\hbar\,\sqrt{(\mathfrak{Re}(\omega_{\overline{n}l}))^{2}+(\mathfrak{Im}(\omega_{\overline{n}l}))^2}
\end{equation}
considering the emission as a coherent superposition of a large number of quantized microstates. In the previous relation the magnitude of the complex frequency $\omega_{nlm}$ is considered, because it represents the "resonant" energy of the transition. 

Neglecting the charge $Q$, considering $\Delta J=-\hbar m$, substituting $\Delta M$ as obtained respectively in \cref{dA} in \cref{d2} and using the quantized area $A$ and $\Delta A$ as defined respectively in \cref{a1,a2}, the following relation \cite{Foit} for the allowed ringdown frequencies can be obtained:
\begin{equation}
\label{freqquant}
|\omega_{\overline{n}lm}|=\frac{c^3}{16\,\pi\,GM}\frac{\overline{n}\alpha\,\sqrt{1-a^2}+8\pi\,am}{\sqrt{1-a^2}},
\end{equation}
which in the case of a Schwarzschild BH reduces to:
\begin{equation}
|\omega_{\overline{n}lm}|=\frac{\alpha\,c^3}{32\pi\,GM}\overline{n}.
\end{equation}

The resulting allowed frequencies computed in the quantized framework are comparable with the frequencies obtained in the context of the classic scenario, since in both cases they are proportional to $1/M$. Now it is possible to numerically evaluate the value of the frequencies obtained due to quantization. Typically the frequency associated to a fundamental QNM is:
\begin{equation}
\omega_{\text{QNM}} = \frac{k\,c^3}{GM},
\end{equation}
with $k$ a dimensionless coefficient.  The ratio between the minimal quantized frequency and the typical ringdown frequency is approximately $12\%$. The effect of quantization imposes a condition on the allowed frequencies, which must be considered alongside the constraints obtained from the QNM solutions in the classical context. The BHAQ hypothesis implies a discrete mass spectrum, which in turn leads to discrete energy transitions corresponding to the mass eigenstates. Compatibility with the emitted energy requires multiple closely spaced degenerate levels for each mass eigenstate, as predicted in ST or LQG. Consequently, this results in a spectrum constructed as a coherent superposition of coherent states, which may differ from the classically predicted spectrum. In this way, the BHAQ hypothesis could lead to variations in the QNM frequencies that are experimentally observable.

\begin{figure}[htbp]
\begin{center}
\includegraphics[scale=0.55]{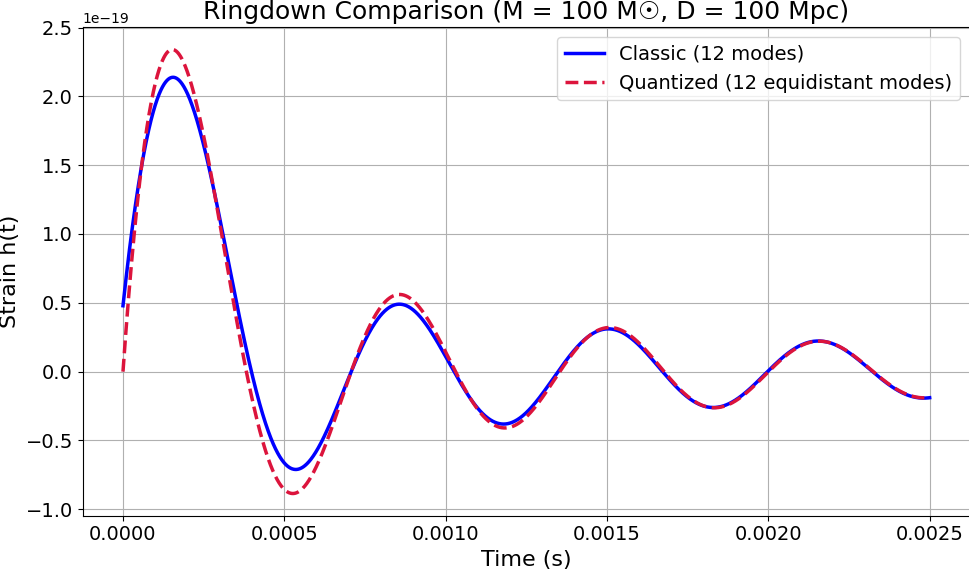}
\caption{Comparison between classical QNMs and quantum-modified QNMs, obtained by either rejecting or modifying the frequencies in accordance with the quantum prediction. The plot uses the first 12 modes.}
\label{figQNMq}
\end{center}
\end{figure}

\begin{figure}[htbp]
\begin{center}
\includegraphics[scale=0.55]{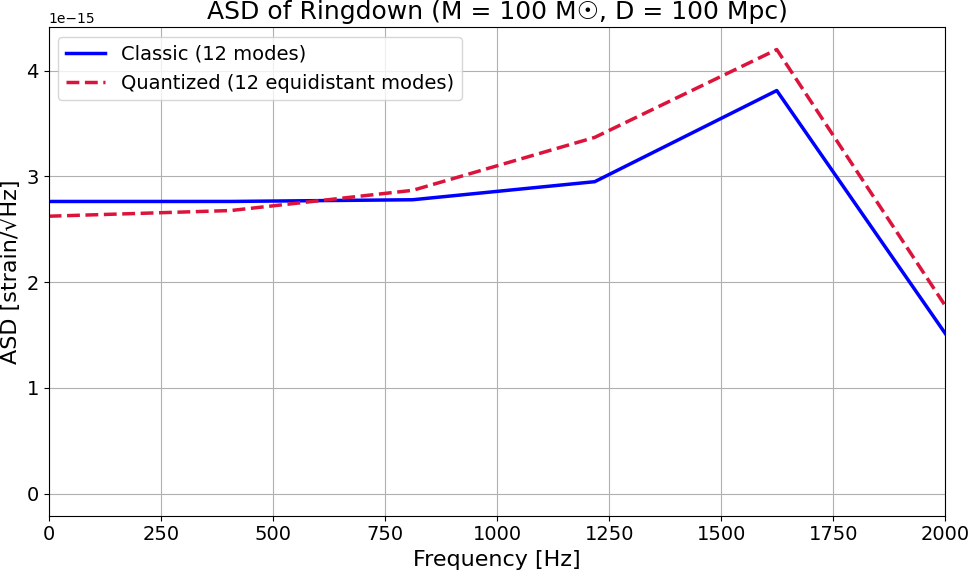}
\caption{Comparison between the spectra calculated as Fourier transform of the expected signals for the classical QNMs and quantum-modified QNMs, obtained by either rejecting or modifying the frequencies in accordance with the quantum prediction. The plot uses the first 12 modes.}
\label{fig4}
\end{center}
\end{figure}

This consideration motivates a simplified illustrative model: a redefinition of the frequencies such that they are spaced proportionally according to \cref{freqquant}. This model is not derived from a specific quantum gravity theory but rather serves as a heuristic representation of possible quantum effects, based on the hypothesis that the QNM frequencies are quantized. This framework illustrates how a quantum-modified discrete emission spectrum could differ from that predicted by classical QNM theory. Moreover, considering the behavior of their complex parts, Maggiore had already hypothesized that QNM frequencies should be quantized \cite{Maggiorequant}. In \cref{figQNMq}, we illustrate how the ringdown phase may be modified by redistributing the frequencies so that they become equally spaced. We evaluate the expected ringdown spectrum for a remnant BH with $M=100M_{\odot}$ at a distance of $100 \text{Mpc}$. We consider the first 12 modes predicted by the most accurate approach to QNMs, namely the continued fraction solution. The plot is limited to these 12 modes, since for higher orders the damping factor becomes dominant and the corresponding modes are strongly suppressed. In \cref{fig4}, we show the Amplitude Spectral Density (ASD) of the foreseen spectrum for the same ringdown emission cases. For a deterministic signal the ASD is essentially the magnitude of its Fourier transform, normalized per unit frequency, that is the distribution of the signal’s amplitude across frequencies: $ASD(f)=\sqrt{2/T}\,|\tilde{h}(f)|$, where $T$ is the signal duration, and $\tilde{h}(f)$ is the Fourier trasform of the strain as a function of the frequency $f$. Also in this case, the differences induced by the possible quantization scenarios considered are visible.

As already illustrated before, any comparison between the measured spectrum and the predictions of classical or quantum-modified models must be performed after correcting the observed frequencies to the source frame using the estimated redshift $z$. All the plots are made using the source frequencies, obtained correcting the observed ones.\footnote{In the case of our plots, at a distance of $100\text{Mpc}$ the redshift value is $z=0.023$, hence the frequency correction is $\sim2\%$.}

\section{Quantum Gravity affecting the BH photon sphere}
\label{sec:QGphotonsphere}

QG models can also be tested in the context of GW emission during the ringdown phase, through the modifications they can introduce on the BH near geometry. This may affect the effective potential related to \cref{Schr} affecting the QNM frequencies or producing echo effects. 

Effective field theory (EFT) approaches to QG introduce higher-curvature corrections or non-local terms \cite{Donoghue}, modifying the gravitational action:
\begin{equation}
S=\int\sqrt{-g}\left[\frac{c^3}{16\pi G}R+l_{Pl}^{2}\,\mathcal{L}_{Pl}+\mathcal{L}_{matt}\right]
\end{equation}
where $\mathcal{L}_{Pl}$ is the Lagrangian density for Planck-scale perturbations, and $\mathcal{L}_{matt}$ is the matter Lagrangian density. We have made the assumption that the only QG perturbation scale is encoded by the Planck length. In this context, the Einstein equations can be expressed as a perturbative series:
\begin{equation}
G_{\mu\nu}+l_{Pl}^{2}\,H_{\mu\nu}+\mathcal{O}(l_{Pl}^{4})=\frac{8\pi\,G}{c^4}T_{\mu\nu}
\end{equation}
As a consequence, one obtains, for the Schwarzschild spacetime metric, a corrected line element of the form:
\begin{equation}
ds^2=-f(r)\,c^2\,dt^{2}+g(r)\,dr^{2}+r^{2}d\theta^{2}+r^{2}\sin^{2}{\theta}d\phi^{2}.
\end{equation}
For simplicity we make the assumption that $g(r)=f^{-1}(r)$, preserving the usual symmetry of BH associated metrics. The metric function $f(r)$ includes QG corrections:
\begin{equation}
f(r) = f_0(r) + \Delta f(r) = 1 - \frac{2\,GM}{c^{2}\,r} + \Delta f(r).
\end{equation}
$\Delta f(r)$ admits a perturbative expansion in the small parameter $l_{Pl}/r$
\begin{equation}
\Delta f(r) =\frac{2\,GM}{c^{2}\,r} \sum_{n=2}^{\infty} \beta_n \left( \frac{l_{Pl}}{r} \right)^n,
\end{equation}
where $\beta_{n}$ are dimensionless coefficients that encode the strength of higher-order quantum corrections. The leading correction typically starts at $n=2$, consistent with the structure of effective field theories of gravity.

A non-perturbative modification capturing similar effects is given by the Bardeen regular BH \cite{Rodrigues}, where $f(r)$ is considered in the form:
\begin{equation}
\label{potpertqg}
f(r) \simeq 1 - \frac{2GM r^2}{c^{2}\,(r^2 + a_0^2)^{3/2}},
\end{equation}
where $a_{0}$ is a [Length]$^3$ dimension constant, which introduces a regularizing behavior at small $r$ \cite{DeLorenzo}.

The general form of the Regge--Wheeler or Zerilli potential is perturbed by the introduction of this modification to the metric. In the case of a perturbed Schwarzschild spacetime, the modified potential can be written as:
\begin{equation}
V(r) = V_{0}(r) \left( 1 + \frac{\Delta f(r)}{f_{0}(r)} \right).
\end{equation}
In this case, the tortoise coordinate $r_{*}$ is defined as:
\begin{equation}
\frac{dr_{*}}{dr} = \frac{1}{f_{0}(r) + \Delta f(r)}.
\end{equation}
Therefore, even if the correction to the potential $V(r)$ is small, the transformation to the tortoise coordinate can amplify its effect, modifying the associated frequencies.  The induced modifications amount to a few percent in the effective potential \cite{Gong} and may affect the predicted QNM frequencies by a similar limited magnitude.

\begin{figure}[htbp]
\begin{center}
\includegraphics[scale=0.55]{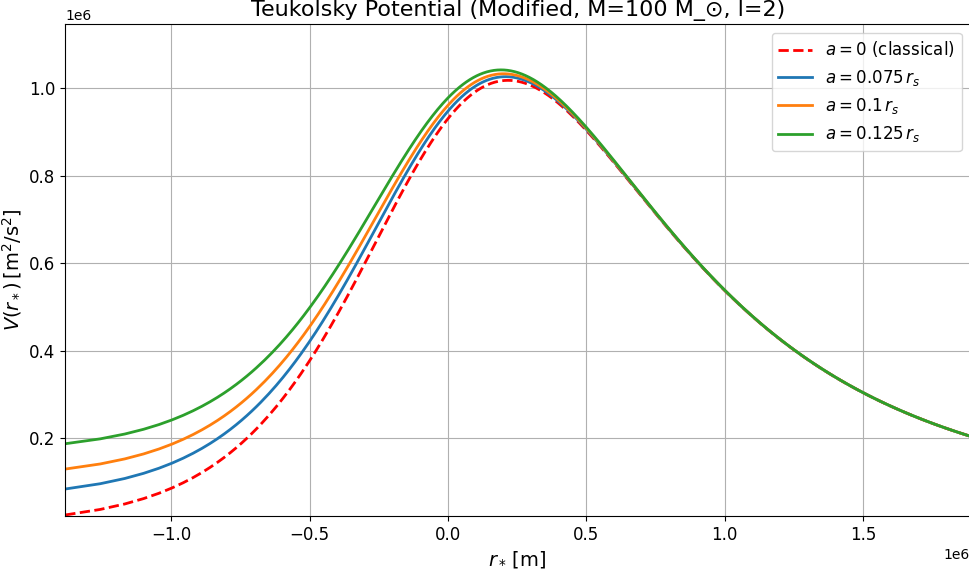} 
\caption{Comparison of the Teukolsky potential with quantum gravity modfications.}
\label{figT}
\end{center}
\end{figure}

In \cref{figT} we report the modifications induced in the Teukolsky potential by quantum gravity perturbations represented in \cref{potpertqg} form. The plot is obtained for a Shwarzschikd BH remnant of $100\,M_\odot$, with $l=2$, considering QG corrections of some percent points of the BH radius, such as foreseen in some BH regularization models \cite{DeLorenzo}.

In some effective models of QG—such as hairy BHs with tensor or scalar hairs \cite{Tattersall}, wormholes and black-bounce metrics \cite{Hadi}, thick-brane extra-dimensional models \cite{Deng}, and toy Bekenstein–Mukhanov models—the predicted quantum corrections \cite{Agullo} also affect the effective potentials appearing in the wave equations for test fields or perturbations. These can be represented as:
\begin{equation}
V_*(r) = V_{\text{eff}}(r) + \Delta V_{Pl}(r),
\end{equation}  
where $\Delta V_{Pl}(r)$ contains perturbative corrections typically of order $(l_{Pl}/r)^{n}$ with $n\geq2$. It should be emphasized that the amplitude of $\Delta V_{Pl}(r)$ is typically very small, so in realistic scenarios the effect on the potential may be extremely subtle. This result, obtained in the context of an effective quantum theory of gravity, suggests that perturbations can introduce an additional peak in the effective potential, enabling the possibility of late-time GW echoes during the ringdown phase, caused by this newly introduced maximum. 

In braneworld-inspired scenarios where the four-dimensional metric is modified by an $r$-dependent warp factor, the effective Schwarzschild line element takes the form:
\begin{equation}
ds^2_{\text{eff}}=e^{2A(r)}\left(-f_{0}(r)\,c^2\,dt^2+f_{0}^{-1}(r)dr^2+r^2d\Omega^2\right).
\end{equation}
The corresponding Regge–Wheeler \cref{effpot1} or Zerilli \cref{effpot2} potentials are given by:
\begin{align}
&V(r)=f_{\text{eff}}(r) \,\ U(r)\\
&f_{\text{eff}}(r)=e^{2A(r)}f(r)=e^{2A(r)}\left(1-\frac{2\,GM}{c^2\,r}\right)\notag
\end{align}
and its extrema satisfy the condition:
\begin{equation}
    V'(r)=f'_{\text{eff}}(r)U(r)+f_{\text{eff}}U'(r)=2A'(r)e^{2A(r)}f(r)U(r)+e^{2A(r)}\big[f'(r)U(r)+f(r)U'(r)\big]=0.
\end{equation}
A similar expression can be found in the case of Teukolsky potential \cref{effpot3}, hence every BH spacetime can be affected by this modification.

In the cases previously analyzed in \cref{effpot1,effpot2,effpot3} the potentials featured a single peak, which reflects incoming perturbations while allowing some to transmit into the BH. This process leads to the initial GW emission, known as the primary ringdown signal.
In the Schwarzschild case this equation admits a single solution associated with the photon sphere, the inclusion of a non-trivial warp profile introduces additional terms proportional to $A'(r)$, which can generate further critical points. In some warp scenarios, therefore, the effective potential can develop a second maximum depending on the contribution of the warp factor $A(r)$. The main peak corresponds to the BH photon sphere and is located outside the EH, whereas the secondary peak is located between the principal maximum and the EH, modifying in an effective way the light ring of the BH. Similar outcomes are found in the contexts of LQG and ST, where quantum modifications to BH EHs are predicted to generate GW echoes \cite{Cardoso,Wang,Barcelo}, even if very weak and not easily detectable. In some models of compact objects the presence of two maxima separated by a minimum forms a cavity that can trap GWs for long periods \cite{Cunha}. This configuration can give rise to long-lived modes, and in certain cases, nonlinear interactions between these modes may cause the trapped energy to grow, potentially destabilizing or even disrupting horizonless objects \cite{Cardoso1}. The effect is often associated with the presence of a stable light ring, which can confine radiation and enhance the probability of nonlinear behavior. However, as discussed in \cite{Hod1}, stable light rings in horizonless spacetimes are typically linked to nonlinear instabilities, whereas BHs —characterized by an absorbing EH— appear to remain dynamically stable configurations. All these models can therefore be tested by searching for late-time echoes in the ringdown phase of GW emission.

\section{Detectability of QNM and quantum-modified QNM frequencies}
\label{sec:detectability}

BHs can be modeled as thermodynamic systems interacting with their surrounding environment. In addition to emitting HR, they can absorb energy from the cosmological background, including photons from the CMB and neutrinos from the C$\nu$B, potentially influencing their mass and thermodynamic evolution. As open thermodynamic systems, BHs may also experience modifications to their ringdown emission spectrum. In particular, the absorption of CMB photons and C$\nu$B neutrinos, which follow a continuous blackbody spectrum, can affect the discrete frequencies of the QNMs, especially those modified by quantum effects. These discrete modes may be broadened, shifted, or partially obscured by the continuous spectrum induced by the BH's interaction with the environment. Detecting quantization effects in the ringdown therefore requires that the discrete QNM frequency spectrum is not overwhelmed by this continuous background. To address this, we can model the ringdown frequencies using again the Lee-Friedrichs formalism.  The QNM frequencies $\Omega_{n}(M)$ depend on the BH mass, which can change due to particle absorption. However, our analysis focuses on the QNMs during the ringdown phase, which typically lasts about $10^{-3}$--$10^{-2}\,\mathrm{s}$. During this short interval, the mass variation induced by the energy of photons or neutrinos falling into the BH is negligible. Specifically, the average energy of the cosmic backgrounds is very small, and the number of absorbed particles is limited due to the brevity of the ringdown. Finally, the hamiltonian \cref{LF} can be amended adding the potential related to the interaction with the cosmic backgrounds.

The introduced potential $V_{CMB}$ governs the interaction between the continuous photon energy spectrum and the discrete predicted QNM frequencies. This potential represents the interaction of the BH with the thermal bath formed by CMB photons. The interaction with the C$\nu$B can be treated in a similar way adding another effective potential for the interaction with the background neutrinos. The $V_{CMB}$ term is obtained as the product of the photon absorption probability times the QNM energy induced modifications:
\begin{align}
\label{V}
&V_{CMB} = \sum_{n}\alpha_{n} \int_0^{+\infty} \Big[ k_{n}(\omega_{\gamma},\,t) \, |\Psi_{n}\rangle\langle\omega_{\gamma}| + \text{h.c.}\Big] \, d\omega_\gamma\\
&k_{n}(\omega_{\gamma},\,t)=\int_{0}^{\Delta t}\hbar \, \Delta \Omega_n(\omega_\gamma,\,t) \sqrt{ \Phi(\omega_\gamma) \, \sigma_{\rm abs}(\omega_\gamma)}\, d\tau, \notag
\end{align}
where $\Delta t$ is the time interval considered for the photon absorption (in the case of ringdown $\Delta t\sim10^{-3}$--$10^{-2}$ s), $\Omega_n$ is the $n$-th discrete QNM frequency, and $\omega_\gamma$ is the photon frequency with a continuous spectrum. $\Phi(\omega_\gamma)$ is the photon flux as a function of frequency:
\begin{equation}
\Phi(\omega_{\gamma}) = \frac{\omega_{\gamma}^{2}}{4\pi^{3}c^{2}} \frac{1}{\exp\left(\hbar \omega_{\gamma}/K_B T\right) - 1},
\end{equation}
$\sigma_{\rm abs}(\omega_{\gamma})$ is the BH radiation absorption cross section:
\begin{equation}
\sigma_{\rm abs} \sim A_{BH} = 4\pi r_{S}^2 = 16\pi\left(\frac{GM}{c^{2}}\right)^2,
\end{equation}
based on the low energy photon cross section for a Schwarzshild BH, appropriate for the low average energy CMB photons. $\Delta \omega_n(\omega_\gamma)$ represents the modification induced on the QNM frequency by the absorption of a photon with frequency $\omega_\gamma$:
\begin{equation}
\Delta\omega_{n}(\omega_{\gamma},\,t) = \frac{\partial \omega_{n}}{\partial M} \frac{\hbar \, \omega_{\gamma}}{c^{2}} \simeq -\frac{\omega_{n}}{M} \frac{\hbar \, \omega_{\gamma}}{c^{2}} = -\alpha_{n} \frac{\hbar \, \omega_{\gamma}}{GM^{2}},
\end{equation}
where the QNM frequency $\omega_{n} = \alpha_{n} c^{3}/GM$ has been used. This is a semi-classical estimate, providing the order of magnitude, assuming linear response and neglecting back-reaction effects. Under this approximation for $\Delta\omega_{n}(\omega_{\gamma})$ the value of $g_{n}(\omega_{\gamma}$ in \cref{V} can be written as:
\begin{equation}
g_{n}(\omega_{\gamma})\simeq \hbar \, \Delta \omega_n(\omega_\gamma,\,t) \sqrt{ \Phi(\omega_\gamma) \, \sigma_{\rm abs}(\omega_\gamma)} \, \Delta t
\end{equation}
The associated resonance amplitude can be computed as:
\begin{equation}
\Gamma_{n}=2\hbar \, \gamma_{n}+\Gamma_{n}^{cont},
\end{equation}
where $\Gamma_{n}^{cont}$ is the decay rate (resonance amplitude)--computed for $\Delta t$ duration of the ringdown $(10^{-2}-10^{-3}\,\text{s})$--generated by the interaction with the continuum spectrum of the CMB, and can be expressed in a form analogous to the Fermi Golden Rule:
\begin{equation}
\Gamma_n^{cont}=2 \, \mathfrak{Im}(\Sigma_{n}(E)) \sim 2 \pi |\Delta \omega_n(\omega_\gamma,\,t)|^{2} \, \Phi(\omega_\gamma) \, \sigma_{\rm abs}(\omega_\gamma)\,\Delta t^{2}.
\end{equation}
$\gamma_n$ is the intrinsic damping of the QNM (due to GW emission), while $\Gamma_{n}^{\rm cont}$ is induced by the interaction with the environment.
The continuum resonance amplitude for all physically observable BHs is completely negligible, well below the threshold of $10^{-65}$ rad/s when evaluated as a frequency amplitude. Since the interactions with the CMB and the C$\nu$B are mediated by terms of equal magnitude, it is possible to state that the discrete structure of the QNM spectrum is not spoiled by these interactions with the backgrounds. Consequently, although a theoretical broadening of the discrete modes caused by interactions with continuous states is expected, it remains practically unobservable. The dominant contribution arises from damping, which produces a visible effect. However, this damping-induced broadening still allows the distinct discrete QNM frequencies to remain observable, as experimentally observed \cite{KAGRA}. As a result, the discrete nature of the QNM spectrum is preserved, and it is conceivable that it also holds for a possible QNM spectrum affected by BHQA. This indicates that different harmonics of the QNMs can be observed during the ringdown phase, potentially allowing the detection of the supposed quantum-induced modifications to the QNM frequencies.

\section{Quantum Gravity in the GW propagation} 
\label{sec:QGpropagation}

Another aspect of GW physics that can be affected by QG concerns possible effects on propagation. Before addressing possible QG modifications to GW propagation, it is important to distinguish them from the standard frequency shifts already predicted by GR and caused by the redshift effect. The cosmological expansion produces a redshift that lowers the observed frequency and stretches the waveform in time by the same factor. This effect is purely kinematical and is routinely corrected in GW data analysis using the luminosity distance and an assumed cosmological model. These “classical” redshift effects must be removed from the data before searching for additional propagation effects, such as dispersion or decoherence, that could signal deviations from GR or the presence of a quantum structure of spacetime.

A common feature of all QG models is the dependence of spacetime geometry on the energy scale being probed. In this context, the presumed quantum structure of spacetime can impact the propagation of GWs by modifying their dispersion relations—that is, by introducing an energy-dependent speed of propagation. 
Starting from the scenario predicted by some QG models, such as for instance Doubly Special Relativity (DSR) and similar models \cite{DSR,HMSR}, the Standard Model Extension (SME) \cite{SME}, LQG \cite{LQG}, certain ST scenarios \cite{ST}, and others effective theories of QG \cite{Cost,WP}, we can write the modified dispersion relations in the form:
\begin{equation}
E^{2}=p^{2}\left(1+\epsilon\left(\frac{E}{E_{Pl}}\right)^{\alpha}\right),
\end{equation}
where $\alpha$ is a model depending exponent, and $E_{Pl}$ is the Planck energy, that is the suppression scale of the perturbation. In this context, considering the foreseen modification to the speed of light 
\begin{equation}
\bar{c}=c\left(1+\frac{\epsilon}{2}\left(\frac{E}{E_{Pl}}\right)^{\alpha}\right), 
\end{equation}
an uncertainty on the probed length can be introduced as:
\begin{equation}
\label{DeltaL}
\Delta L\sim L^{1-\alpha}l_{Pl}^{\alpha},
\end{equation}
where $\alpha=1$ for the DSR scenario, $\alpha=2/3$ in the case of holography model, and $\alpha=1/2$ for the random walk model \cite{Ng}. It is then possible to evaluate, within each QG scenario, the perturbations induced in the GW strain as a consequence of these distance fluctuations:
\begin{equation}
\delta h\simeq\frac{\Delta L}{L}
\end{equation}
In the case of DSR theories, where $\Delta L=l_{Pl}$, the resulting strain perturbation is extremely small. For an interferometer such as the Einstein Telescope (ET) \cite{ET_2} or Cosmic Explorer (CE) \cite{Cosmic_Explorer}, with proposed arms of approximately $10$ km, the strain is constrained to a contribution $\delta h\lesssim 10^{-39}$ far below the observable threshold of $h\sim10^{-22}\div10^{-23}$.

In the holographic scenario, the dominant contribution arises from: $\Delta L=L^{1/3}l_{Pl}^{2/3}$, leading to $\delta h\sim10^{-26}$ for ET—several orders of magnitude below current or near-future sensitivity. While this value remains unobservable at present, future advances in experimental technology could make it accessible.

The stochastic (random walk) model predicts a distance fluctuation $\Delta L=\sqrt{L\cdot l_{Pl}}$ resulting in a raw strain perturbation of $\delta h\sim10^{-20}$. However, this represents the maximum theoretical deviation before accounting for statistical averaging, and then filtering, or comparison with instrumental sensitivity. Because high-frequency stochastic fluctuations tend to average out, this effect is strongly suppressed and effectively undetectable with current instruments.

Energy-dependent variations in GW propagation length can also induce decoherence in the wave packet. The resulting decoherence can be described by the phase perturbation:
\begin{equation}
\Delta\phi=\frac{1}{\hbar}\int{mc^{2}\,\delta g^{00}\,dt}\simeq\frac{2\pi\Delta L}{\lambda},
\end{equation}
where $\lambda$ is the wavelength of the propagating perturbation. This result is obtained assuming that $\Delta L$ is the distance fluctuation and $\delta g^{00}$ is the related spacetime metric fluctuation.
If $\Delta\phi$ is stochastic, the decoherence arises from dephasing within the wave packet and can be formalized by introducing the residual coherence factor $C$:
\begin{equation}
C(\Delta\phi)=\exp\left({-\frac{1}{2}\langle(\Delta \phi)^{2}\rangle}\right) \end{equation}
and introducing it in the definition of the metric perturbation:  
\begin{equation}
\label{deco}
h_{**}(t,\,x)\rightarrow h_{**}C(\Delta\phi)=h_{**}(t,\,x)\exp\left({-\frac{1}{2}\langle(\Delta \phi)^{2}\rangle}\right).
\end{equation}
Among all considered scenarios, only the holographic framework predicts potentially detectable decoherence. In this case, a decoherence factor of  $(1-C)\sim10^{-4}$ is expected—currently beyond detection limits, but possibly accessible with future instruments.

CPT violation can induce similar effects in the GW propagation. The violation of CPT symmetry can be interpreted as the presence of a background tensor field in the spacetime, as hypothesized in the SME \cite{SME}. This background induces coherent asymmetries in the metric fluctuations, which are not purely stochastic. The associated scaling is typically $\alpha=1$, corresponding to coherent but anisotropic fluctuations. At the macroscopic level, this results in a direction-dependent, polarization-sensitive propagation of gravitational waves, i.e., a birefringent effect. The predictions of the SME are more complex because, in this model, the underlying diffeomorphism symmetry of GR is violated (rather than modified), leading to a broader phenomenology \cite{Wang2}. In the context of the SME, CPT-violating perturbations can be considered and investigated \cite{Kostelecky2}. In this framework, the gravitational action is written by adding operators that violate Lorentz invariance to the usual Lagrangian. The simplest CPT-violating term is of the form:
\begin{equation}
\mathcal{L}_{\rm CPT} \sim (k_V)_\kappa \, \epsilon^{\kappa\lambda\mu\nu} \, h_{\lambda\alpha} \, \partial_\mu \, h_\nu{}^\alpha.
\end{equation}
The resulting modification of the dispersion relations can be written as:
\begin{equation}
\omega_\pm^2=|\mathbf{k}|^2\pm(\kappa_V\cdot\mathbf{k})\, |\mathbf{k}|^2
\end{equation}
where $\omega_{\pm}$ are related to the right- and left-handed polarizations of the GW, defined as $h_{R}=1/\sqrt{2}(h_{+}-ih_{\times})$ and $h_{L}=1/\sqrt{2}(h_{+}+ih_{\times})$. The different propagation velocities of the two polarizations can induce a birefringence effect. The accumulated phase difference after a propagation $L$ is:
\begin{equation}
\Delta\Phi=\int_0^L (\omega_{+}-\omega_{-})\,dt\sim(\kappa_{V}\cdot\hat{\mathbf{k}})\,L.
\end{equation}
The LIV parameter $\kappa_{V}$ can be constrained using the relation:
\begin{equation}
(\kappa_{V})_{(min)}=\frac{\Delta\Phi_{min}}{L}.
\end{equation}
It is expected that the ET detector will be able to detect a minimum phase difference of $\Delta\Phi_{min}\sim10^{-3}\,\text{rad}$ \cite{ET_3}. Therefore, it will be possible to place constraints on the order of $\kappa_{V}\sim10^{-32}\,\text{m}^{-1}$, improving the actual constraint of $2-3$ orders of magnitude. 

As a final remark it is possible to state that the effect of QG in propagation alter dispersion relations and thus affect GW propagation \cite{Calcagni}, introducing decoherence or birefringence. 

Identifying QG effects in the ringdown phase of BH mergers requires separating contributions from the quantum structure of BHs from those arising due to QG-induced modifications in GW propagation. Moreover, modified GW propagation can help constrain QG models that predict different dispersion relations for GWs and other messengers, such as electromagnetic signals. For instance, searching for time delays between the arrival of GWs and $\gamma$-rays from the same astrophysical event can provide valuable insight into this domain \cite{Calcagni}.

\section{Experimental Perspectives}
\label{label:experimental}

The main question is if during the next decades of actual century we will be able to detect a QG signature in GW emission. We have already shown that the ideal sector concerns the detection of GWs emitted during the ringdown phase. The possibility of detecting the presumed quantization effects of BHs in the GW emission from ringdown is solely linked to the sensitivity expected from the experimental apparatus. New projects as ET \cite{ET_2}  and CE \cite{Cosmic_Explorer} have the target to increase the amplitude sensitive of a factor 10 even expanding the detector bandwidth well below 10 Hz.
CE is conceived, at least in the first phase, as a detector based on the LIGO technology and the sensitivity improvement is mainly due to the factor 10 increase of the arm length of the interferometer. 
ET takes a different approach: the order-of-magnitude jump in sensitivity is achieved through a combination of new technologies, together with an increase in the detector arm lengths, which is moderate compared to CE. ET will operate at cryogenic temperatures (about 10 K) to reduce the thermal vibrations of the suspended optics, and the test masses will be heavier than 200 kg to reduce noise due to radiation pressure.
A xylophone configuration, consisting of two co-located interferometers, will optimally cover two different frequency ranges. The low-frequency interferometer (ET-LF) will use heavier test masses to reduce the radiation pressure component of quantum noise and will be cooled to low temperatures to overcome the thermal noise limitation of the mirror suspensions. The high-frequency interferometer (ET-HF) will store several MW of light power in the main optical cavities to reduce the shot-noise limitation.
In the low-frequency regime (5–50 Hz), two fundamental noise sources compete: thermal noise of the suspensions and seismic noise from ground vibrations. To mitigate disturbances at very low frequencies ($\sim$10 Hz) due to residual seismic noise and Newtonian noise, the ET interferometers will be located in an underground site.

\begin{figure}[htbp]
\begin{center}
\includegraphics[scale=0.55]{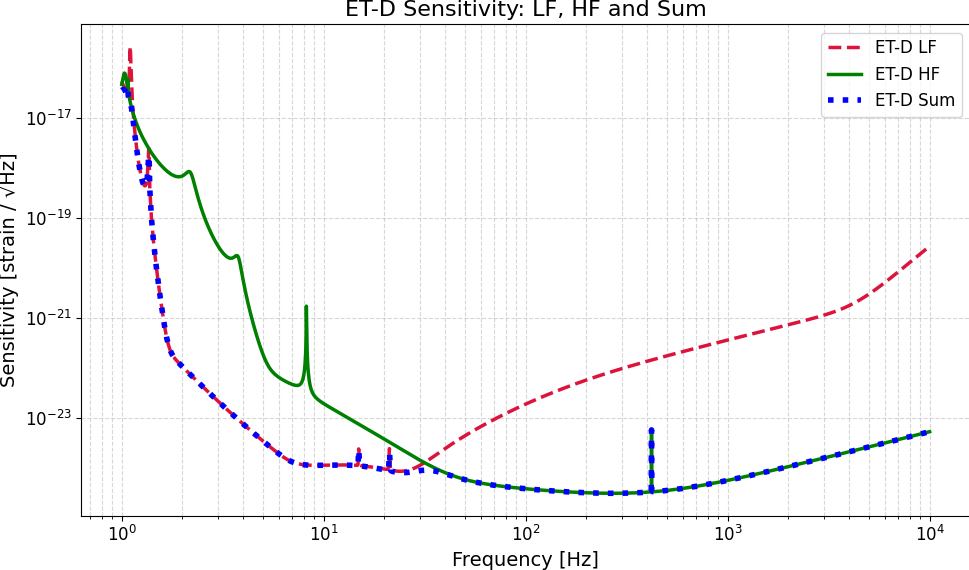}
\caption{Sensitivity curve expected for the ET detector - Low Frequency detector - High Frequency detector - resulting sensitivity curve as sum of the prevoius plots \cite{ET_sensitivity}.}
\label{fig6}
\end{center}
\end{figure}

The predicted performance of ET results from the combination of ET-LF and ET-HF sensitivity curves. 
The crossover frequency of the sensitivities of the ET-LF and ET-HF is at about 35 Hz. Above this frequency the high-frequency interferometer is limited only by two noise sources: mirror thermal-noise limits the sensitivity between 40 and 200 Hz, while the high-frequency section is limited by quantum optics noise.
The final curve  depends of few hundreds parameters characterizing the performances of the various sub-systems of this complex apparatus. As an example it include auxiliary long optical cavities to squeeze the e.m. vacuum of the two interferometers. Their function is to rotate the phase of the vacuum squeezed ellipse along the frequency bandwidth of the detector \cite{ET_3}. 
Here we refer to the sensitivity curves public available on the ET site, named ET-LF-D, ET-HF-D and ET-D \cite{ET_sensitivity} shown in the \cref{fig6}.

\begin{figure}[htbp]
\begin{center}
\includegraphics[scale=0.55]{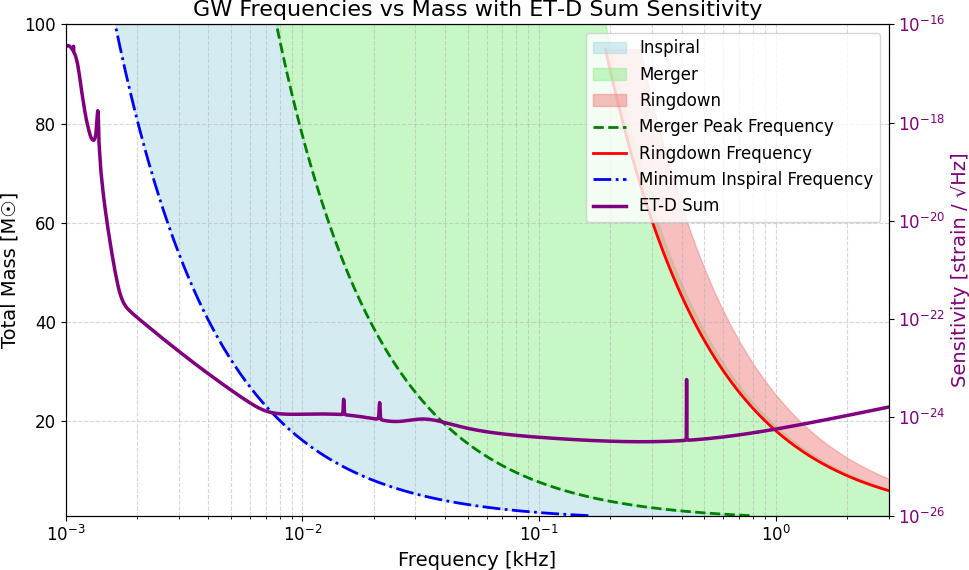}
\caption{Comparison of the expected frequencies with the projected sensitivity curve of the ET. The frequencies are calculated for coalescing binary systems with equal masses.}
\label{fig7}
\end{center}
\end{figure}

In \cref{fig7}, we show a comparison between the expected frequencies for the three phases of GW emission and the predicted signal-to-noise ratio curve for the ET. The frequencies are computed for binary coalescing systems composed of objects with equal masses. Recall that the proportional relationship between the mass of the coalescing objects and the gravitational wave emission frequency is $\omega\sim1/M$. The expected frequencies are represented as colored bands to indicate that they tend to vary over time, increasing for instance as the coalescing objects approach. The different frequency bands are separated by the minimum frequency lines for each distinct phase. In the same plot is reported the sensitivity curve for the ET detector, obtained from the data in reference \cite{ET}. On the left vertical axis are shown the values of the total mass of the coalescing system in units of solar masses $M_{\odot}$, while on the right vertical axis are shown the values of the signal-to-noise ratio referred to the sensitivity function, defined as:
\begin{equation}
h_{n}(f)=\sqrt{S_{n}(f)}.
\end{equation}
The emission frequencies are indicated on the abscissa axis.

Now, through the Signal-to-Noise Ratio ($SNR$) function, it becomes possible to obtain the relative precision, which can be estimated to first approximation as:
\begin{equation}
\frac{\delta h}{h}=\frac{1}{SNR}\frac{1}{\sqrt{T\,f}},
\end{equation}
where $\delta h$ is the uncertainty on the strain $h$ measure, $1/\sqrt{T\,f}\simeq1/N$ and $N$ is the number of complete oscillations observed. in the case of ringdown $N\lesssim10$, since this emission is strongly suppressed by the dumping factor and for an interferometer, such as the ET detector, the expected ratio for a detectable signal is $\delta h/h\sim 10^{-2}\div10^{-3}$. This relative error on the strain can be compared to the sensitivity on the detected frequencies. In fact, from the relation $\delta h/h\sim1/\sqrt{f}$ one can estimate through error propagation that $\delta f/f\sim2\delta h/h$. However, it must be considered that during the analysis of the ringdown phase, the strain is significantly suppressed due to damping, and therefore the proportionality between the strain error and the frequency error changes during the observation of multiple oscillations. Nevertheless, one can conclude that the error should remain sufficiently small to allow for the detection of the hypothesized quantization effects in the frequencies of the ringdown phase.

\begin{figure}[htbp]
\begin{center}
\includegraphics[scale=0.55]{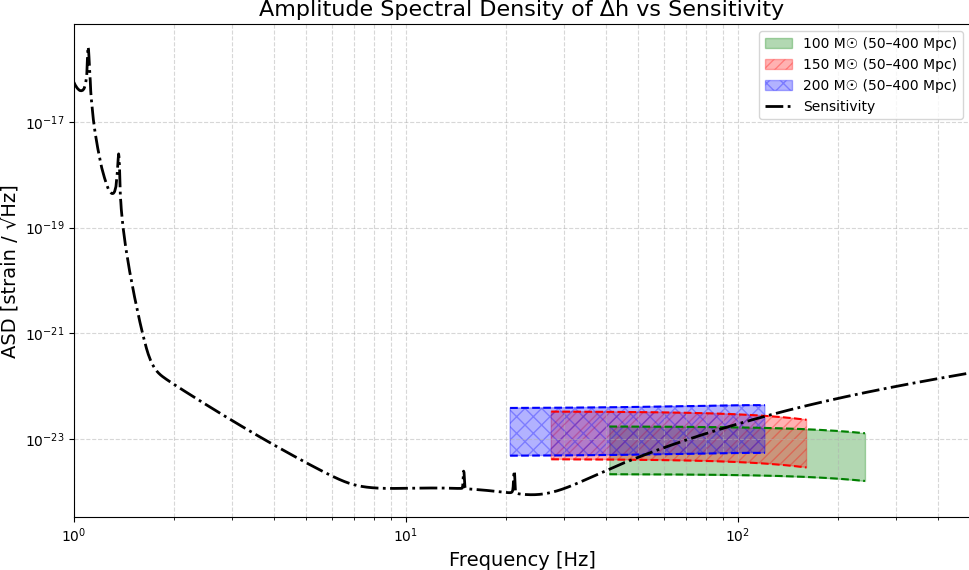} \\[0.5cm]
\includegraphics[scale=0.55]{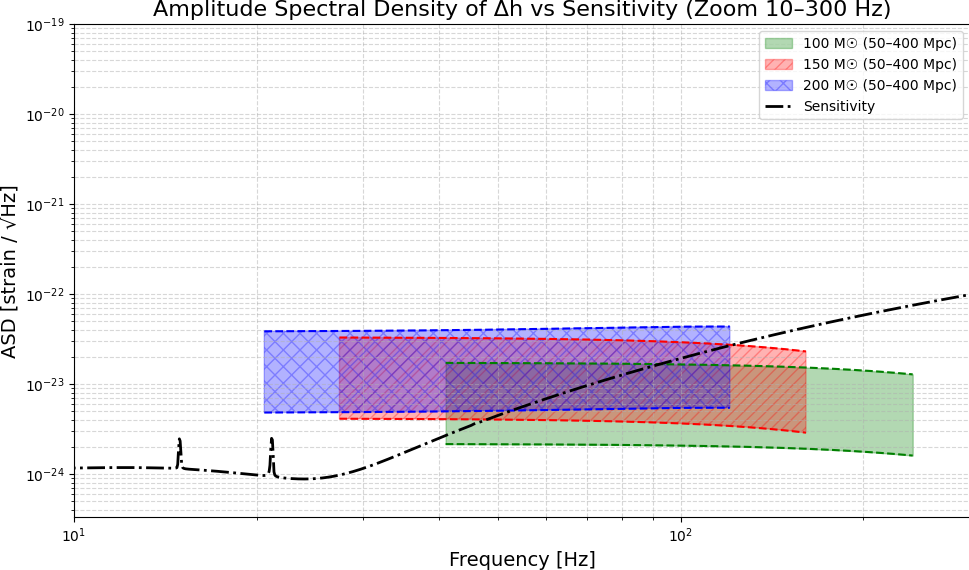}
\caption{Comparison of the ADS of the strain of the classical QNM ringdown and the quantized scenario with the expected ET sensitivity curve, with the strain evaluated for remnant BHs of 100, 150, and 200 $M_{\odot}$ at distances between 40 and 400 Mpc.}
\label{fig8}
\end{center}
\end{figure}

In \cref{fig8} we present the ADS function of the strain computed as the difference between the strain obtained using classically calculated QNM frequencies and that derived from quantized frequencies according to the quantization model considered in \cref{sec:emission}. This difference is evaluated for the ringdown of three remnant BHs with masses of 100, 150, and 200 $M_{\odot}$. The bands representing these differences are shown for source distances in the range 50–400 Mpc. The ADS is plotted as a function of frequency, enabling direct comparison with the sensitivity curve of ET. It is observed that, as the mass increases and the distance decreases, the difference between the classical strain and the strain from the quantized model lies above the sensitivity curve. For suitable combinations of mass and distance, the difference between the classical and quantized models is expected to exceed the uncertainty imposed by the detector sensitivity. In conclusion, with the next generation of detectors and under ideal conditions, it may be possible to test the hypothesis of quantization of BH energy levels.

The evaluation of the QG impact on propagation of GWs is important for determining its effects on detector sensitivity and the possibility of observing quantum signatures in BH structure. By applying the perturbation strain as obtained in \cref{deco}, it is possible to compute the predicted correction caused by decoherence foreseen in various scenarios of QG. For typical propagation distances, the decoherence contributions predicted by the DSR scenario, the holographic model, and the random walk model are negligible. For example, in the case of the holographic model -which produces the largest contribution- for propagation distances on the order of tens or hundreds of Mpc, the ratio $\delta h/h<10^{-4}$ for decoherence effects is expected—currently beyond detection limits, but possibly accessible with future instruments. The DSR case is independent of the source distance and yields a ratio $\delta h/h\sim10^{-70}$, while the random walk model does depend on distance but remains far below the holographic case. Decoherence effects, therefore, are relatively limited and, depending on the model considered, may only introduce a completely negligible background noise, undetectable by current instruments. The modification of the GW propagation speed can also cause a deformation of the expected spectrum, but this effect appears to be subdominant compared to other expected perturbations caused, for instance, by the presumed quantization of BHs. The only way to verify whether a presumed quantization of spacetime can affect the propagation of GWs is, therefore, to compare their propagation speed with that of other cosmic messengers that might be affected in a different way by QG.

\section{Conclusion}

In recent years, the direct detection of GWs has opened a new frontier in fundamental physics, enabling tests of GR in the previously inaccessible strong-field regime. This advancement has allowed for the experimental study of BHs, potentially providing insights into the physics governing them \cite{population,Verymassive,LIGO}. The recent observation of QNMs, with the ability to resolve different overtones \cite{KAGRA}, has indeed shown that experimental advances could make the hypothesized quantization-induced modifications in GW emission accessible.  
In this work, we illustrated how the emission of GWs during the ringdown phase can be related to the loss of mass from BHs. The structure of BHs can be connected to quantum mechanics by considering the relationship between HR, BH entropy, and the proposed quantization of BH mass and EH area. This connection arises in the context of models such as LQG, ST, and the Bekenstein--Mukhanov proposal \cite{Bekenstein3,Hod}, all of which suggest discrete spectra for BH mass and area, therefore, the BH mass eigenstates may be quantized. Quantum-induced modifications to the expected frequencies are predicted to scale as $1/M$, similar to the level spacing predicted in the classical GW scenario. In particular, the fact that the QNM frequencies of the ringdown are of the same order of magnitude as those in the classical scenario, provides a potential observational window into QG effects. Considering how the ringdown emission from a BH remnant could be modified by the hypothesized quantization of mass eigenstates, we have investigated different scenarios. At first we examined the Bekenstein’s original model, that cannot account for the total energy emitted during BH merger as GWs. Indeed, in this scenario, the emission of a single graviton for a quantized jump between mass eigenstates cannot carry the entire energy of the emission. We have therefore shown that Bekenstein-like emission could occur as a collective emission of coherent graviton packets, behavior compatible with that of a Bose-Einstein condensate quantum superfluid \cite{Manikandan,Hennigar} made up of multiple coherent states \cite{Dvali,Dvali2}. This scenario is possible introducing the degeneracy predicted in ST \cite{Strominger} or LQG \cite{Meissner,Sahlmann}, admitting a large number of degenerate entropy states with similar energy. Treating BHs within the semiclassical Lee--Friedrichs framework \cite{Xiao,Lonigro}, we provided an estimate of the possibility of resolving the different QNM modes, even under quantized BH hypothesis. In this scenario, the emission of QNMs with discrete frequencies, as observed experimentally \cite{KAGRA}, is preserved. Quantization could ultimately modify the predicted frequencies, introducing quantized spacing between the QNMs, creating a testable observable singature. 

We further investigated whether the coupling of BHs with their surrounding environment can affect the detectability of QNMs and the predicted quantum-induced perturbations on the ringdown. We treated BHs as open thermodynamic systems interacting with their environment, modeling the ringdown phase through an effective Hamiltonian derived from the Lee--Friedrichs model. We showed that this coupling, while it may broaden the QNM eigenvalues, should not smear the predicted discrete oscillation modes into a continuum, thus potentially allowing the detection of quantum perturbations.  

Next, we considered how various proposed QG scenarios could influence the propagation of GWs. The hypothesized quantized structure of spacetime can affect GW propagation by inducing decoherence effects caused by metric fluctuations \cite{Ng}. In most of the scenarios considered, these effects are negligible and therefore do not interfere with the observation of the proposed quantized structure of BHs via the ringdown emission phase. Only with next-generation detectors will some QG frameworks become accessible and testable, particularly the holographic scenario, and CPT-odd perturbations as predicted by some models such as the SME.

Finally, we discussed the potential to explore this domain with future interferometers \cite{ET_2,Cosmic_Explorer}, such as the ET, assessing the experimental feasibility of this line of research. With the right combination of physical properties of the observed merging BHs—namely, as the remnant BH mass increases and the distance decreases—it may be possible, under ideal experimental conditions, to put the BHAQ hypothesis to the test in the future.

\section*{Acknowledgments}

The authors would like to express their sincere gratitude to Stefano Liberati, Vittorio Gorini and Sergio Cacciatori for their valuable advice, which greatly helped to improve this research work.



\begin{thebibliography}{99}

\bibitem{KAGRA}
A.~G.~Abac \textit{et al.} [KAGRA, Virgo and LIGO Scientific],
``GW250114: Testing Hawking{\textquoteright}s Area Law and the Kerr Nature of Black Holes,''
Phys. Rev. Lett. \textbf{135} (2025) no.11, 111403
doi:10.1103/kw5g-d732
[arXiv:2509.08054 [gr-qc]].

\bibitem{Bekenstein}
J.~D.~Bekenstein,
``Black holes and the second law,''
Lett. Nuovo Cim. \textbf{4} (1972), 737-740
doi:10.1007/BF02757029

\bibitem{Bekenstein2}
J.~D.~Bekenstein,
``Black holes and entropy,''
Phys. Rev. D \textbf{7} (1973), 2333-2346
doi:10.1103/PhysRevD.7.2333

\bibitem{Meissner}
K.~A.~Meissner,
``Black hole entropy in loop quantum gravity,''
Class. Quant. Grav. \textbf{21} (2004), 5245-5252
doi:10.1088/0264-9381/21/22/015
[arXiv:gr-qc/0407052 [gr-qc]].

\bibitem{Sahlmann}
H.~Sahlmann,
``Toward explaining black hole entropy quantization in loop quantum gravity,''
Phys. Rev. D \textbf{76} (2007), 104050
doi:10.1103/PhysRevD.76.104050
[arXiv:0709.2433 [gr-qc]].

\bibitem{Bekensteinbound}
J.~D.~Bekenstein,
``A Universal Upper Bound on the Entropy to Energy Ratio for Bounded Systems,''
Phys. Rev. D \textbf{23} (1981), 287
doi:10.1103/PhysRevD.23.287

\bibitem{Berti2}
E.~Berti,
``Black hole quasinormal modes: Hints of quantum gravity?,''
Conf. Proc. C \textbf{0405132} (2004), 145-186
[arXiv:gr-qc/0411025 [gr-qc]].

\bibitem{Carneiro}
S.~Carneiro and C.~Pigozzo,
``Quasinormal modes and horizon area quantisation in Loop Quantum Gravity,''
Gen. Rel. Grav. \textbf{54} (2022), 20
[erratum: Gen. Rel. Grav. \textbf{54} (2022) no.3, 25]
doi:10.1007/s10714-022-02910-x
[arXiv:2012.00227 [gr-qc]].

\bibitem{Maldacena}
J.~M.~Maldacena,
``Black holes in string theory,''
[arXiv:hep-th/9607235 [hep-th]].

\bibitem{Bekenstein3}
J.~D.~Bekenstein and V.~F.~Mukhanov,
``Spectroscopy of the quantum black hole,''
Phys. Lett. B \textbf{360} (1995), 7-12
doi:10.1016/0370-2693(95)01148-J
[arXiv:gr-qc/9505012 [gr-qc]].

\bibitem{Hod}
S.~Hod,
``Bohr's correspondence principle and the area spectrum of quantum black holes,''
Phys. Rev. Lett. \textbf{81} (1998), 4293
doi:10.1103/PhysRevLett.81.4293
[arXiv:gr-qc/9812002 [gr-qc]].

\bibitem{Strominger}
A.~Strominger and C.~Vafa,
``Microscopic origin of the Bekenstein-Hawking entropy,''
Phys. Lett. B \textbf{379} (1996), 99-104
doi:10.1016/0370-2693(96)00345-0
[arXiv:hep-th/9601029 [hep-th]].

\bibitem{Cost}
A.~Addazi, J.~Alvarez-Muniz, R.~Alves Batista, G.~Amelino-Camelia, V.~Antonelli, M.~Arzano, M.~Asorey, J.~L.~Atteia, S.~Bahamonde and F.~Bajardi, \textit{et al.}
``Quantum gravity phenomenology at the dawn of the multi-messenger era{\textemdash}A review,''
Prog. Part. Nucl. Phys. \textbf{125} (2022), 103948
doi:10.1016/j.ppnp.2022.103948
[arXiv:2111.05659 [hep-ph]].

\bibitem{WP}
R.~Alves Batista, G.~Amelino-Camelia, D.~Boncioli, J.~M.~Carmona, A.~di Matteo, G.~Gubitosi, I.~Lobo, N.~E.~Mavromatos, C.~Pfeifer and D.~Rubiera-Garcia, \textit{et al.}
``White paper and roadmap for quantum gravity phenomenology in the multi-messenger era,''
Class. Quant. Grav. \textbf{42} (2025) no.3, 032001
doi:10.1088/1361-6382/ad605a
[arXiv:2312.00409 [gr-qc]].

\bibitem{Foit}
V.~F.~Foit and M.~Kleban,
``Testing Quantum Black Holes with Gravitational Waves,''
Class. Quant. Grav. \textbf{36} (2019) no.3, 035006
doi:10.1088/1361-6382/aafcba
[arXiv:1611.07009 [hep-th]].

\bibitem{Coates}
A.~Coates, S.~H.~V{\"o}lkel and K.~D.~Kokkotas,
``On black hole area quantization and echoes,''
Class. Quant. Grav. \textbf{39} (2022) no.4, 045007
doi:10.1088/1361-6382/ac4618
[arXiv:2201.03245 [gr-qc]].

\bibitem{ET_2}
M.~Punturo, M.~Abernathy, F.~Acernese, B.~Allen, N.~Andersson, K.~Arun, F.~Barone, B.~Barr, M.~Barsuglia and M.~Beker, \textit{et al.}
``The Einstein Telescope: A third-generation gravitational wave observatory,''
Class. Quant. Grav. \textbf{27} (2010), 194002
doi:10.1088/0264-9381/27/19/194002

\bibitem{Cosmic_Explorer} 
M.~Evans, A.~Corsi, C.~Afle, A.~Ananyeva, K.~G.~Arun, S.~Ballmer, A.~Bandopadhyay, L.~Barsotti, M.~Baryakhtar and E.~Berger, \textit{et al.}
``Cosmic Explorer: A Submission to the NSF MPSAC ngGW Subcommittee,''
[arXiv:2306.13745 [astro-ph.IM]].

\bibitem{Damour}
T.~Damour and A.~Nagar,
``A new analytic representation of the ringdown waveform of coalescing spinning black hole binaries,''
Phys. Rev. D \textbf{90} (2014) no.2, 024054
doi:10.1103/PhysRevD.90.024054
[arXiv:1406.0401 [gr-qc]].
       
\bibitem{Berti}
E.~Berti, V.~Cardoso and A.~O.~Starinets,
``Quasinormal modes of black holes and black branes,''
Class. Quant. Grav. \textbf{26} (2009), 163001
doi:10.1088/0264-9381/26/16/163001
[arXiv:0905.2975 [gr-qc]].

\bibitem{Nollert}
H.~P.~Nollert,
``TOPICAL REVIEW: Quasinormal modes: the characteristic `sound' of black holes and neutron stars,''
Class. Quant. Grav. \textbf{16} (1999), R159-R216
doi:10.1088/0264-9381/16/12/201

\bibitem{Regge}
T.~Regge and J.~A.~Wheeler,
``Stability of a Schwarzschild singularity,''
Phys. Rev. \textbf{108} (1957), 1063-1069
doi:10.1103/PhysRev.108.1063

\bibitem{Zerilli}
F.~J.~Zerilli,
``Effective potential for even parity Regge-Wheeler gravitational perturbation equations,''
Phys. Rev. Lett. \textbf{24} (1970), 737-738
doi:10.1103/PhysRevLett.24.737

\bibitem{Moncrief}
V.~Moncrief,
``Gravitational perturbations of spherically symmetric systems. I. The exterior problem.,''
Annals Phys. \textbf{88} (1974), 323-342
doi:10.1016/0003-4916(74)90173-0

\bibitem{Martel}
K.~Martel and E.~Poisson,
``Gravitational perturbations of the Schwarzschild spacetime: A Practical covariant and gauge-invariant formalism,''
Phys. Rev. D \textbf{71} (2005), 104003
doi:10.1103/PhysRevD.71.104003
[arXiv:gr-qc/0502028 [gr-qc]].

\bibitem{Teukolsky}
S.~A.~Teukolsky,
``Perturbations of a rotating black hole. 1. Fundamental equations for gravitational electromagnetic and neutrino field perturbations,''
Astrophys. J. \textbf{185} (1973), 635-647
doi:10.1086/152444

\bibitem{Arbey}
A.~Arbey, J.~Auffinger, M.~Geiller, E.~R.~Livine and F.~Sartini,
``Hawking radiation by spherically-symmetric static black holes for all spins: Teukolsky equations and potentials,''
Phys. Rev. D \textbf{103} (2021) no.10, 104010
doi:10.1103/PhysRevD.103.104010
[arXiv:2101.02951 [gr-qc]].

\bibitem{Sasaki}
M.~Sasaki and T.~Nakamura,
``A Class of New Perturbation Equations for the Kerr Geometry,''
Phys. Lett. A \textbf{89} (1982), 68-70
doi:10.1016/0375-9601(82)90507-2

\bibitem{Sasaki2}
M.~Sasaki and T.~Nakamura,
``Gravitational Radiation From a Kerr Black Hole. 1. Formulation and a Method for Numerical Analysis,''
Prog. Theor. Phys. \textbf{67} (1982), 1788
doi:10.1143/PTP.67.1788

\bibitem{Schutz}
B.~F.~Schutz and C.~M.~Will,
``BLACK HOLE NORMAL MODES: A SEMIANALYTIC APPROACH,''
Astrophys. J. Lett. \textbf{291} (1985), L33-L36
doi:10.1086/184453

\bibitem{Iyer}
S.~Iyer and C.~M.~Will,
``BLACK HOLE NORMAL MODES: A SEMIANALYTIC APPROACH. 1. FOUNDATIONS,''
Print-86-0935 (WASH.U.,ST.LOUIS).

\bibitem{Leaver}
E.~W.~Leaver,
``An Analytic representation for the quasi normal modes of Kerr black holes,''
Proc. Roy. Soc. Lond. A \textbf{402} (1985), 285-298
doi:10.1098/rspa.1985.0119

\bibitem{Ferrari}
V.~Ferrari and B.~Mashhoon,
``New approach to the quasinormal modes of a black hole,''
Phys. Rev. D \textbf{30} (1984), 295-304
doi:10.1103/PhysRevD.30.295

\bibitem{Hawking3}
S.~W.~Hawking,
``Particle Creation by Black Holes,''
Commun. Math. Phys. \textbf{43} (1975), 199-220
[erratum: Commun. Math. Phys. \textbf{46} (1976), 206]
doi:10.1007/BF02345020

\bibitem{Hawking4}
S.~W.~Hawking,
``Black Holes and Thermodynamics,''
Phys. Rev. D \textbf{13} (1976), 191-197
doi:10.1103/PhysRevD.13.191

\bibitem{Polchinski}
J.~Polchinski,
``The black hole information problem.,''
doi:10.1142/9789813149441{\_}0006
[arXiv:1609.04036 [hep-th]].

\bibitem{Maldacena1}
J.~M.~Maldacena,
``The Large $N$ limit of superconformal field theories and supergravity,''
Adv. Theor. Math. Phys. \textbf{2} (1998), 231-252
doi:10.4310/ATMP.1998.v2.n2.a1
[arXiv:hep-th/9711200 [hep-th]].

\bibitem{Xiao}
Z.~Xiao and Z.~Y.~Zhou,
``On the generalized Friedrichs-Lee model with multiple discrete and continuous states,''
[arXiv:2310.14937 [hep-ph]].

\bibitem{Lonigro}
D.~Lonigro,
``The self-energy of Friedrichs{\textendash}Lee models and its application to bound states and resonances,''
Eur. Phys. J. Plus \textbf{137} (2022) no.4, 492
doi:10.1140/epjp/s13360-022-02690-y
[arXiv:2109.02939 [math-ph]].

\bibitem{ETH}
V.~Balasubramanian, A.~Lawrence, J.~M.~Magan and M.~Sasieta,
``Microscopic Origin of the Entropy of Astrophysical Black Holes,''
Phys. Rev. Lett. \textbf{132} (2024) no.14, 141501
doi:10.1103/PhysRevLett.132.141501
[arXiv:2212.08623 [hep-th]].

\bibitem{Manikandan}
S.~K.~Manikandan and A.~N.~Jordan,
``Bosons falling into a black hole: A superfluid analogue,''
Phys. Rev. D \textbf{98} (2018) no.12, 124043
doi:10.1103/PhysRevD.98.124043
[arXiv:1811.03209 [gr-qc]].

\bibitem{Hennigar}
R.~A.~Hennigar, R.~B.~Mann and E.~Tjoa,
``Superfluid Black Holes,''
Phys. Rev. Lett. \textbf{118} (2017) no.2, 021301
doi:10.1103/PhysRevLett.118.021301
[arXiv:1609.02564 [hep-th]].

\bibitem{Dvali}
G.~Dvali and C.~Gomez,
``Black Hole's Quantum N-Portrait,''
Fortsch. Phys. \textbf{61} (2013), 742-767
doi:10.1002/prop.201300001
[arXiv:1112.3359 [hep-th]].

\bibitem{Dvali2}
G.~Dvali and C.~Gomez,
``Landau{\textendash}Ginzburg limit of black hole's quantum portrait: Self-similarity and critical exponent,''
Phys. Lett. B \textbf{716} (2012), 240-242
doi:10.1016/j.physletb.2012.08.019
[arXiv:1203.3372 [hep-th]].

\bibitem{Maggiorequant}
M.~Maggiore,
``The Physical interpretation of the spectrum of black hole quasinormal modes,''
Phys. Rev. Lett. \textbf{100} (2008), 141301
doi:10.1103/PhysRevLett.100.141301
[arXiv:0711.3145 [gr-qc]].

\bibitem{Donoghue}
J.~F.~Donoghue,
``General relativity as an effective field theory: The leading quantum corrections,''
Phys. Rev. D \textbf{50} (1994), 3874-3888
doi:10.1103/PhysRevD.50.3874
[arXiv:gr-qc/9405057 [gr-qc]].

\bibitem{Rodrigues}
M.~E.~Rodrigues and M.~V.~de Sousa Silva,
``Bardeen Regular Black Hole With an Electric Source,''
JCAP \textbf{06} (2018), 025
doi:10.1088/1475-7516/2018/06/025
[arXiv:1802.05095 [gr-qc]].

\bibitem{DeLorenzo}
T.~De Lorenzo, C.~Pacilio, C.~Rovelli and S.~Speziale,
``On the Effective Metric of a Planck Star,''
Gen. Rel. Grav. \textbf{47} (2015) no.4, 41
doi:10.1007/s10714-015-1882-8
[arXiv:1412.6015 [gr-qc]].

\bibitem{Gong}
H.~Gong, S.~Li, D.~Zhang, G.~Fu and J.~P.~Wu,
``Quasinormal modes of quantum-corrected black holes,''
Phys. Rev. D \textbf{110} (2024) no.4, 044040
doi:10.1103/PhysRevD.110.044040
[arXiv:2312.17639 [gr-qc]].

\bibitem{Tattersall}
O.~J.~Tattersall,
``Quasi-Normal Modes of Hairy Scalar Tensor Black Holes: Odd Parity,''
Class. Quant. Grav. \textbf{37} (2020) no.11, 115007
doi:10.1088/1361-6382/ab839b
[arXiv:1911.07593 [gr-qc]].

\bibitem{Hadi}
H.~Hadi and R.~Naderi,
``Gravitational memory effects of black bounces and a traversable wormhole,''
Eur. Phys. J. C \textbf{84} (2024) no.4, 343
doi:10.1140/epjc/s10052-024-12718-5
[arXiv:2402.09435 [gr-qc]].

\bibitem{Deng}
W.~Deng, S.~Long, Q.~Tan, Z.~C.~Chen and J.~Jing,
``Scalar-gravitational quasinormal modes and echoes in a five dimensional thick brane,''
[arXiv:2508.20937 [gr-qc]].

\bibitem{Agullo}
I.~Agullo, V.~Cardoso, A.~D.~Rio, M.~Maggiore and J.~Pullin,
``Potential Gravitational Wave Signatures of Quantum Gravity,''
Phys. Rev. Lett. \textbf{126} (2021) no.4, 041302
doi:10.1103/PhysRevLett.126.041302
[arXiv:2007.13761 [gr-qc]].

\bibitem{Cunha}
P.~V.~P.~Cunha, E.~Berti and C.~A.~R.~Herdeiro,
``Light-Ring Stability for Ultracompact Objects,''
Phys. Rev. Lett. \textbf{119} (2017) no.25, 251102
doi:10.1103/PhysRevLett.119.251102
[arXiv:1708.04211 [gr-qc]].

\bibitem{Cardoso1}
V.~Cardoso and P.~Pani,
Nature Astron. \textbf{1} (2017) no.9, 586-591
doi:10.1038/s41550-017-0225-y
[arXiv:1709.01525 [gr-qc]].

\bibitem{Hod1}
S.~Hod,
``On the number of light rings in curved spacetimes of ultra-compact objects,''
Phys. Lett. B \textbf{776} (2018), 1-4
doi:10.1016/j.physletb.2017.11.021
[arXiv:1710.00836 [gr-qc]].

\bibitem{Cardoso}
V.~Cardoso, V.~F.~Foit and M.~Kleban,
``Gravitational wave echoes from black hole area quantization,''
JCAP \textbf{08} (2019), 006
doi:10.1088/1475-7516/2019/08/006
[arXiv:1902.10164 [hep-th]].

\bibitem{Wang}
Q.~Wang, N.~Oshita and N.~Afshordi,
``Echoes from Quantum Black Holes,''
Phys. Rev. D \textbf{101} (2020) no.2, 024031
doi:10.1103/PhysRevD.101.024031
[arXiv:1905.00446 [gr-qc]].

\bibitem{Barcelo}
C.~Barcel{\'o}, R.~Carballo-Rubio and L.~J.~Garay,
``Gravitational wave echoes from macroscopic quantum gravity effects,''
JHEP \textbf{05} (2017), 054
doi:10.1007/JHEP05(2017)054
[arXiv:1701.09156 [gr-qc]].

\bibitem{DSR}
G.~Amelino-Camelia, J.~Kowalski-Glikman, G.~Mandanici and A.~Procaccini,
``Phenomenology of doubly special relativity,''
Int. J. Mod. Phys. A \textbf{20} (2005), 6007-6038
doi:10.1142/S0217751X05028569
[arXiv:gr-qc/0312124 [gr-qc]].

\bibitem{HMSR}
M.~D.~C.~Torri, V.~Antonelli and L.~Miramonti,
``Homogeneously Modified Special relativity (HMSR): A new possible way to introduce an isotropic Lorentz invariance violation in particle standard model,''
Eur. Phys. J. C \textbf{79} (2019) no.9, 808
doi:10.1140/epjc/s10052-019-7301-7
[arXiv:1906.05595 [hep-th]].

\bibitem{SME}
D.~Colladay and V.~A.~Kostelecky,
``Lorentz violating extension of the standard model,''
Phys. Rev. D \textbf{58} (1998), 116002
doi:10.1103/PhysRevD.58.116002
[arXiv:hep-ph/9809521 [hep-ph]].

\bibitem{LQG}
M.~Bojowald and G.~M.~Hossain,
``Loop quantum gravity corrections to gravitational wave dispersion,''
Phys. Rev. D \textbf{77} (2008), 023508
doi:10.1103/PhysRevD.77.023508
[arXiv:0709.2365 [gr-qc]].

\bibitem{ST}
G.~Amelino-Camelia, M.~Arzano, Y.~Ling and G.~Mandanici,
Class. Quant. Grav. \textbf{23} (2006), 2585-2606
doi:10.1088/0264-9381/23/7/022
[arXiv:gr-qc/0506110 [gr-qc]].

\bibitem{Ng}
Y.~J.~Ng and E.~S.~Perlman,
``Probing Spacetime Foam with Extragalactic Sources of High-Energy Photons,''
Universe \textbf{8} (2022) no.7, 382
doi:10.3390/universe8070382
[arXiv:2205.12852 [gr-qc]].

\bibitem{Wang2}
Q.~Wang, J.~M.~Yan, T.~Zhu and W.~Zhao,
``Modified gravitational wave propagations in linearized gravity with Lorentz and diffeomorphism violations and their gravitational wave constraints,''
Phys. Rev. D \textbf{111} (2025) no.8, 084064
doi:10.1103/PhysRevD.111.084064
[arXiv:2501.11956 [gr-qc]].

\bibitem{Kostelecky2}
V.~A.~Kosteleck{\'y} and M.~Mewes,
``Testing local Lorentz invariance with gravitational waves,''
Phys. Lett. B \textbf{757} (2016), 510-514
doi:10.1016/j.physletb.2016.04.040
[arXiv:1602.04782 [gr-qc]].

\bibitem{Calcagni}
G.~Calcagni,
``Quantum gravity and gravitational-wave astronomy,''
doi:10.1007/978-981-15-4702-7{\_}30-1
[arXiv:2012.08251 [gr-qc]].

\bibitem{ET_3}
ET~Steering~Committee~Editorial~Team,  
``2020 Design Report Update for the Einstein Telescope,"
[Tech. Rep. [ET-0007B-20]]

\bibitem{ET_sensitivity}
[ET-sensivity], 
https://www.et-gw.eu/index.php/etsensitivities

\bibitem{ET}
M.~Maggiore \textit{et al.} [ET],
``Science Case for the Einstein Telescope,''
JCAP \textbf{03} (2020), 050
doi:10.1088/1475-7516/2020/03/050
[arXiv:1912.02622 [astro-ph.CO]].

\bibitem{population}
R. Abbott et al. (LIGO Scientific Collaboration, Virgo Collaboration, and KAGRA Collaboration), 
``Population of merging compact binaries inferred using gravitational waves through GWTC-3" 
Phys. Rev. X testbf{13} (2023) 011048.

\bibitem{Verymassive}
LIGO Scientific Collaboration, Virgo Collaboration, and KAGRA Collaboration
``GW231123: a Binary Black Hole Merger with Total Mass 190-265" 
$M_{\odot}$,
arXiv:2507.08219; https://arxiv.org/abs/2507.08219.

\bibitem{LIGO}
B.~P.~Abbott \textit{et al.} [LIGO Scientific and Virgo],
``Observation of Gravitational Waves from a Binary Black Hole Merger,''
Phys. Rev. Lett. \textbf{116} (2016) no.6, 061102
doi:10.1103/PhysRevLett.116.061102
[arXiv:1602.03837 [gr-qc]].

\end{thebibliography}


\end{document}